\documentclass[10pt,conference,]{IEEEtran}
\IEEEoverridecommandlockouts

\usepackage[bibbreaks=tight,
            paragraphs=tight,
            floats=normal,
            mathspacing=normal,
            wordspacing=normal,
            tracking=normal,
            bibnotes=tight,
            charwidths=normal,
            mathdisplays=normal,
            leading=normal,
            indent=normal,
            lists=normal,
            bibliography=normal,
            title=normal,
            sections=normal,
            margins=normal,
            ]{savetrees}

\usepackage[english]{babel}
\usepackage[utf8x]{inputenc}
\usepackage[T1]{fontenc}
\usepackage[colorinlistoftodos]{todonotes}
\usepackage{cite}

\usepackage{amsmath,amssymb,mathrsfs,bbm,amsthm}
\usepackage{graphicx}
\usepackage[colorinlistoftodos]{todonotes}
\usepackage[colorlinks=true, allcolors=blue]{hyperref}
\usepackage{multirow}
\usepackage[lined,linesnumbered,ruled]{algorithm2e}
\usepackage{mathrsfs}
\usepackage{xfrac}

\newcommand{\cp}[1]{\ifmmode {\mathcal{#1}}\else ${\mathcal{#1}}$\fi}

\newcommand{\bD}{\boldsymbol{D}}

\newcommand{\bE}{\boldsymbol{E}}

\newcommand{\bI}{\boldsymbol{I}}

\newcommand{\bL}{\boldsymbol{L}}

\newcommand{\bQ}{\boldsymbol{Q}}
\newcommand{\bR}{\boldsymbol{R}}

\newcommand{\bV}{\boldsymbol{V}}
\newcommand{\bW}{\boldsymbol{W}}

\newcommand{\bY}{\boldsymbol{Y}}

\newcommand{\bc}{\boldsymbol{c}}
\newcommand{\bd}{\boldsymbol{d}}

\newcommand{\br}{\boldsymbol{r}}
\newcommand{\by}{\boldsymbol{y}}

\newcommand{\bu}{\boldsymbol{u}}

\newcommand{\bx}{\boldsymbol{x}}

\newcommand{\bq}{\boldsymbol{q}}

\newcommand{\calE}{\mathcal{E}}
\newcommand{\calG}{\mathcal{G}}
\newcommand{\calN}{\mathcal{N}}

\newcommand{\calV}{\mathcal{V}}

\newcommand{\bdelta}{\boldsymbol{\delta}}

\newcommand{\bDelta}{\boldsymbol{\Delta}}

\newcommand{\cb}[1]{\boldsymbol{#1}}

\newcommand{\Ex}{\mathbb{E}}
\newcommand{\vect}{\operatorname{vec}}

\usepackage{color}  

\definecolor{darkgreen}{rgb}{0.0, 0.85, 0.0}

\begin{document}

\title{Online Graph-Based Change Point Detection \\ in Multiband Image Sequences 
\thanks{The work of R. Borsoi and J. Bermudez was funded in part by CNPq under grants 304250/2017-1, 409044/2018-0, 141271/2017-5 and 204991/2018-8.}
\thanks{The work of C. Richard was funded in part by the ANR under grant ANR-19-CE48-0002, and by the 3IA C\^ote d'Azur Senior Chair program.}}



\author{R.~A.~Borsoi$^{\,*,\,\mathsection}$,~~ C.~Richard$^{\,*}$,~~ A.~Ferrari$^{\,*}$,~~ J.~Chen$^{\,\dagger}$,~~ J.~C.~M.~Bermudez$^{\,\mathsection,\,\ddagger}$ \\[6pt]
 \small $^{\star}$ Universit\'{e} C\^{o}te d'Azur, CNRS, OCA, Nice, France \\
 \small $^{\mathsection}$ Federal University of Santa Catarina, Florian\'opolis, Brazil \\
 \small $^{\dagger}$ Northwestern Polytechnical University, Xi'an, China\\
 \small $^{\ddagger}$ Catholic University of Pelotas, Pelotas, Brazil
}

\maketitle

\begin{abstract}

The automatic detection of changes or anomalies between multispectral and hyperspectral images collected at different time instants is an active and challenging research topic. To effectively perform change-point detection in multitemporal images, it is important to devise techniques that are computationally efficient for processing large datasets, and that do not require knowledge about the nature of the changes. In this paper, we introduce a novel online framework for detecting changes in multitemporal remote sensing images. Acting on neighboring spectra as adjacent vertices in a graph, this algorithm focuses on anomalies concurrently activating groups of vertices corresponding to compact, well-connected and spectrally homogeneous image regions. It fully benefits from recent advances in graph signal processing to exploit the characteristics of the data that lie on irregular supports. Moreover, the graph is estimated directly from the images using superpixel decomposition algorithms. The learning algorithm is scalable in the sense that it is efficient and spatially distributed. Experiments illustrate the detection and localization performance of the method.
\end{abstract}


\section{Introduction}

The increasing availability of multitemporal multi- and hyperspectral devices allows for a detailed analysis of the evolution of a scene over time. This can potentially benefit different applications, ranging from agricultural and forestry monitoring, natural disaster and urban landscape analysis~\cite{liu2019reviewChangeDetectionGRSM}, to surveillance and other industry \mbox{problems~\cite{banerjee2009hyperspectralVideoIlluminationInvTracking,stefansson2020hyperspectralVideoDryingWood}.}
Among these applications, the detection of changes or anomalies between images acquired at different time instants is an active and challenging research topic~\cite{jianya2008reviewRemoteSensingCD,liu2019reviewChangeDetectionGRSM}.
%
%
%
%
%
The earliest change detection (CD) works consisted of post-classification algorithms, which compare supervised classification results of the two images~\cite{jianya2008reviewRemoteSensingCD,liu2019reviewChangeDetectionGRSM}. Although conceptually simple, those methods depend on training data, whose availability is often limited.
This motivated the development of unsupervised techniques, which usually consists in analysing some kind of distance between feature-based representations of the pixels in both images~\cite{liu2019reviewChangeDetectionGRSM}. 
Other popular approaches search for a transformation of the data to a lower-dimensional feature space that highlights the changes (e.g., multivariate alteration detection (MAD), temporal PCA, or unmixing)~\cite{liu2019reviewChangeDetectionGRSM}.

Traditionally, most CD algorithms were focused on processing a single pair of images (i.e., the bitemporal problem). However, those methods fail to explore the information contained in longer data streams, which has recently become the source of attention. 
For instance,~\cite{bertoluzza2017loopsConsistencyCD} proposed to improve the performance of binary CD by constraining detection results to be consistent across all closed loops of subsequences sampled from the observed time series. 
%
CD decision rules known a priori have also been applied to new pairs of images using domain adaptation techniques~\cite{zanotta2015domainAdaptationDecisionRulesCD}.
In~\cite{vaduva2012latentMetricTimeSeries}, different features were computed in local spatial windows centered at each pixel, for the whole time series. These temporal vectors of features were then classified using methods such as, e.g., K-means.
More applied work often considers precise season-trend parametric models over, e.g., vegetation indices and detect changes by examining the model parameters~\cite{verbesselt2012vegetationCDtimeSeries}.

Other algorithms were devised to operate online, such as~\cite{salmon2013changeDetectionKalmanCovariance}, which detects changes on vegetation index data by monitoring the state covariance matrix of the extended Kalman filter. Another approach used density ratio estimation (learned from training data) to detect changes~\cite{anees2015supervisedDensityRatioNDVItimeSeries}.
In \cite{grobler2012pagesCUSUMmodis} a cumulative sum test is used to detect change points in time series data. However, the pre- and post-change densities have to be estimated a priori.
Recent work also considered multitemporal unmixing to separate the observed images into time series of the material signatures (endmembers) and their fractional abundances for each pixel~\cite{thouvenin2016online,borsoi2020multitemporalUKalmanEM_arxiv}. This can be used to perform CD in the lower dimensional space~\cite{liu2019reviewChangeDetectionGRSM}.
Other methods based on, e.g., total-variation~\cite{frecon2016onlineCDTotalVariation} and low-rank and sparse decomposition~\cite{candes2011robustPCA} can also be considered to perform CD. However, reconciling a low computational complexity with flexibility to incorporate a priori knowledge about the nature of the changes can be difficult in these techniques.


Despite the significant interest in this problem, there is still a need for algorithms that operate online and explore the temporal and the intrinsic structure of the data, while also being unsupervised and computationally efficient.
In this work, we propose an online framework for detecting changes in multitemporal multiband images. We assume that changes occur concurrently in groups of connected, spectrally homogeneous pixels, which are then represented as adjacent vertices in a graph constructed from the superpixel representation of the images.
Instead of considering graph regularizations~\cite{yokoya2017multisensorCoupledUnmixing} or graph-based representations of the images learned from the data~\cite{heas2005trajectoriesDynamicClustersTimeSeries}, which lead to computationally inefficient and offline solutions, the proposed method fully benefits from recent advances in graph signal processing to better exploit the characteristics of the data that lie on irregular supports~\cite{shuman2013signalProcessingGraphsReview}. Moreover, differently from previous bitemporal CD approaches that use superpixels to compute features to classification algorithms~\cite{gong2017superpixelFeaturesCD} or as an ad hoc approach to introduce spatial homogeneity in CD algorithms~\cite{wu2012superpixelCDmajVoting}, we use the superpixel decomposition to generate a graph directly from the observed images. This allows us to exploit theoretically-grounded tools from graph signal processing. The learning algorithm is computationally efficient and spatially distributed, and benefits from temporal information by using a simple strategy that propagates two statistics with different learning rates. 
Simulations illustrate both the detection and localization performance of the proposed method.


\section{Problem Formulation}

We consider a sequence of multi- or hyperspectral images $\bY_{\!t}\in\mathbb{R}^{L\times N}$ with $N$ pixels and $L$ bands, for $t=1,\ldots,T$.
The unsupervised binary CD problem can be formulated as follows: given the image sequence $\bY_{\!t}$, find $\bc_t\in\mathbb{R}^N$, for $t=2,\ldots,T$. where $c_{t,j}=1$ if there occurred a change in pixel $j$ between instants $t-1$ and $t$, and  $c_{t,j}=0$ otherwise.
More precisely, we consider the measurements to be given by:
\begin{align} \label{eq:im_model_i}
    \bY_{\!t} = \bQ_t + \bE_t \,,
\end{align}
where $\bQ_t$ denotes contextually relevant image content, and $\bE_t$ represents additive noise and perturbations that are non-informative changes, which include, e.g., noise, illumination and other acquisition variations~\cite{borsoi2020variabilityReview,Borsoi_2018_Fusion}. This allows us to define the CD problem equivalently as:
\begin{align}
    c_{t,j} = 1 \quad \Longleftrightarrow \quad \bq_{t,j} \neq \bq_{t-1,j}  \,, \quad t=2, \dots, T,
\end{align}
where $\bq_{t,j}$ is the $j^{\rm th}$ column of $\bQ_t$, for $t=1,\ldots,T$.

In this work, we make some simplifying assumptions that facilitate the definition of the change detection problem:

{\bf A1:} Changes occur concurrently in sets of pixels that share semantically similar information (such as buildings, roads, crop fields, etc.). Mathematically, we assume that every pixel $i$ for which $c_{t,i}=1$ is part of one among a set of compact, connected, spectrally homogeneous groups of pixels $\mathscr{C}(k)$, $k=1,\ldots,N_{\mathscr{C}}$ for which $c_{t,j}=1$ for all pixels $j$ in the same group as $i$.

{\bf A2:} Only a single change occurs in a given streaming signal $\by_{t,j}$ of length $T$ ($1\leq t\leq T$).

A1 is reasonable as changes usually occur in compact regions of the image composed of several pixels. A2 allows us to explore temporal information contained in longer sequences~\cite{liu2019reviewChangeDetectionGRSM}, differently from multi- and hyperspectral image CD algorithms that focus in the bitemporal problem (i.e., when $T=2$). Moreover, very long sequences where multiple changes occur can be divided into subsequences of smaller length $T$ in which A2 is satisfied, as long as multiple abrupt changes do not happen in close succession.

A1 and A2 motivate a reformulation of the CD problem as the detection of abrupt variations in the data sequence. This simplified model is more accurate as the ratio of the sampling period of the discrete image sequence to the time required for the occurrence of a significant change increases. We use this simplified model for ease of exposition.

The change point detection (CPD) problem consists of finding $t_c\in\mathbb{N}$ such that
\begin{align}
\begin{split}
    \bq_{t,j} &= \overline{\bq}_{j}, \hspace{8.8ex} t < t_c \,,
    \\
    \bq_{t,j} &= \overline{\bq}_{j} + \bdelta_j, \hspace{4ex} t \geq t_c \,.
\end{split} \label{eq:cpd_model_ii}
\end{align}
where vector $\overline{\bq}_{j}\in\mathbb{R}^L$ is the $j^{\rm th}$ column of matrix $\overline{\bQ}$, which captures the static underlying image content, and $\bdelta_j\in\mathbb{R}^L$ is the $j^{\rm th}$ column of matrix $\bDelta$, which captures  the changes. 
%
This formulation is adequate for different applications such as surveillance or other industry applications~\cite{banerjee2009hyperspectralVideoIlluminationInvTracking,stefansson2020hyperspectralVideoDryingWood}, which can benefit from computationally efficient online algorithms.

\subsection{Superpixel-based graph construction} \label{sec:sppx_graph_constr}

An important aspect of this problem is how groups of pixels $\mathscr{C}(k)$ should be defined. Fundamentally, $\mathscr{C}(k)$ should contain pixels that share semantically similar information (e.g., belonging to the same objects or regions that may experience changes concurrently). We focus at linking pixels that are spatially adjacent and spectrally similar, attempting to respect sharp image borders that might indicate different objects or image regions.
Recently, different works proposed to divide multiband images into disjoint, compact regions by means of image (over)-segmentation methods such as superpixels, ultrametric contour maps, or binary partition trees~\cite{achanta2012slicPAMI,arbelaez2010ultrametricContourMaps,veganzones2014hyperspectralSegmentationBPT}. In particular, superpixels are able to group spectrally similar pixels in compact spatial neighborhoods of average size $S^2$ with excellent preservation of image borders~\cite{achanta2012slicPAMI}. The superpixel decomposition has been sucesfully applied to construct multiscale transformations used in spectral unmixing applications~\cite{borsoi2018superpixels1_sparseU,Borsoi_multiscaleVar_2018,borsoi2019BMUAN}.
In this paper, we consider the superpixel decomposition in order to divide the pixels $n=1,\ldots,N$ into sets $\mathscr{C}(k)$ that group the pixels belonging to each superpixel of the image.

\section{Proposed strategy}

Traditional CD techniques operate individually on each pixel of an image. 
In this work we propose a framework that is able to explore, in a theoretically-grounded manner, the information that changes occur in groups of semantically similar pixels in possibly long image sequences, while at the same time benefiting from temporal information to operate online and computationally efficiently. We base our work on~\cite{ferrari2019distributed}, which we extend to consider multidimensional data, and use a graph constructed from the superpixel decomposition.


We assume that the images are defined over an undirected graph $\calG=\{\calV,\calE,\bW\}$ with $N$ vertices $\calV=\{1,\ldots,N\}$, $M$ edges $(i,j)\in\calE\subseteq\calV\times\calV$ and a (symmetric) adjacency matrix $\bW\in\mathbb{R}^{N\times N}$ whose elements are nonnegative and specify the similarity between the different pixels in the image. We also denote by $\bL\in\mathbb{R}^{N\times N}$ the normalized graph Laplacian of~$\calG$.
According to the discussion in Section~\ref{sec:sppx_graph_constr}, $\calG$ is constructed such that $(i,j)\in\calE$ if and only if pixels $i$ and $j$ belong to the same superpixel, \mbox{in which case $[\bW]_{(i,j)}=1$.}

This definition makes $\mathscr{C}(k)$ equivalent to clusters in $\calG$, and the CD problem becomes closely related to testing for changes in clusters of pixels in the graph. A change will be characterized by a $\bdelta_j\neq\cb{0}$ for~$j$ in a well-connected cluster of vertices of~$\calG$.
%
A detection strategy that is closely related to this problem is the graph Fourier scan statistic (GFSS)~\cite{sharpnack2015graphCD_GFSS}. The GFSS addresses a counterpart of this problem for scalar (one-dimensional) graph signals $\bx\in\mathbb{R}^N$ measured over $\calG$ by testing whether the average value of $\bx$ is constant on all vertices, or if there is a cluster $\mathscr{C}$ of vertices with a different average value~\cite{sharpnack2015graphCD_GFSS}. The GFSS considers the following test statistic:
\begin{align} 
    r_{\rm centrGFSS} = \big\| g\big( \bx \big) \big\|_2 \,,
    \label{eq:stat_GFSS_centralized}
\end{align}
where $g(\bx)$ is the graph-filtered version of $\bx$, where the graph filter $g$ is defined as
\begin{align}
    g\big( \bx \big) = \sum_{n=2}^N h(\mu_n) \big( \bu_n^\top \bx \big) \bu_n \,,
    \label{eq:graph_filter_1}
\end{align}
with $\bu_n\in\mathbb{R}^N$ being the $n^{\rm th}$ eigenvector of $\bL$, $\mu_n$ the eigenvalue associated with $\bu_n$, and
\begin{align}
    h(\mu) = \min\Big\{1,\sqrt{\gamma/\mu} \Big\}\,, \,\, \mu>0 \,.
    \label{eq:graph_filter_2}
\end{align}
Parameter $\gamma>0$ in \eqref{eq:graph_filter_2} controls the behavior of the graph filter that is applied to the vertex-wise test statistic, which attenuates high-frequency variations in $\bx$ while enhancing signals that are consistent across well connected clusters of~$\calG$ (i.e., a low-pass graph filter)~\cite{shuman2013signalProcessingGraphsReview}.
Although the graph filter in~\eqref{eq:graph_filter_1} requires the eigendecomposition of $\bL$, which may be computationally costly, several strategies have been proposed to address this issue. Those typically involve approximating $g$ by a finite impulse response~\cite{segarra2017distributedFIRGraphFilter}, or by autoregressive moving average~\cite{isufi2016ARMA_graphFiltering} graph filters that can be implemented efficiently and in a distributed manner (even for large $N$).

The GFSS, however, cannot be directly applied to the present problem since the underlying hypotheses are not satisfied (i.e., the graph signal is not constant over all vertices under the null hypothesis, and the nodal observations are vector valued pixels). 
If the underlying trend $\overline{\bQ}$ were known, it could be subtracted from $\bY_{\!t}$, and each band of the resulting signals could be tested using the GFSS. Since $\overline{\bQ}$ is not known, we adopt the adaptive strategy first considered in \cite{keriven2020newma,ferrari2019distributed}, where two adaptive filters with different time constants $\lambda$ and $\Lambda$, $0<\lambda<\Lambda<1$, are used according to:
\begin{align} \label{eq:newma_adap}
\begin{split}
    \bV_{\!t} &= (1-\lambda) \bV_{\!t-1} + \lambda \bY_{\!t} \,,
    \\
    \bV_{\!t}' &= (1-\Lambda) \bV_{\!t-1}' + \Lambda \bY_{\!t} \,.
\end{split}
\end{align}
Both $\bV_{\!t}$ and $\bV_{\!t}'$ are weighted averages of $\bY_{\!t}$, which can mitigate the influence of additive noise $\bE_t$ in~\eqref{eq:im_model_i}. Moreover, $\bV_{\!t}'$ captures the short-term variations in $\bY_{\!t}$, from which the long-term average $\bV_{\!t}$ can be subtracted~\cite{keriven2020newma} as:
%
%
\begin{align} \label{eq:differece_img_newma}
    \bD_t = \bV_{\!t}'-\bV_{\!t} \,,
\end{align}
where $\bD_t^\top=[\bd_t^{(1)},\ldots,\bd_t^{(L)}]$. Assuming model~\eqref{eq:cpd_model_ii}, we have that $\bD_t$ is asymptotically zero-mean in the absence of a change-point (i.e., $t<t_c$). On the other hand, after a change-point occurs (i.e., $t\geq t_c$), $\bD_t$ will be non-zero-mean at the clusters affected by the change in $\bDelta$, since $\bV_{\!t}'$ will adapt more quickly than $\bV_{\!t}$ to represent the changes in $\bY_{\!t}$ due to the difference between the time constants of both filters in~\eqref{eq:newma_adap} (i.e., a change can be detected by verifying how the short-term average differs from the long-term trend). 

The graph signal $\bD_t$ is, however, vector valued (at each pixel), what still hinders the use of the GFSS test statistic as defined in~\eqref{eq:stat_GFSS_centralized}. 
Moreover, as noted in~\cite{ferrari2019distributed}, the global test statistic $r_{\rm centrGFSS}$ does not allow one to know in which cluster of $\calG$ the change occurred. An alternative strategy proposed in~\cite{ferrari2019distributed} is to consider the coherent sum of the graph-filtered difference signal at the neighborhood of each vertex of $\calG$, which localizes the changes in the graph.
Since we are testing for changes affecting all bands of pixels in clusters $\mathscr{C}(k)$ (i.e., we do not localize the changes at the band level), we propose a distributed test statistic for each vertex of $\calG$ by summing the squared norms of the coherent sums of $g(\bd_t^{(\ell)})$ over all bands $\ell=1,\ldots,L$. This leads to the following test statistic for each pixel:
\begin{align}
    \br_{\rm daGFSS}(n) = \sum_{\ell=1}^L \bigg(\sum_{m\in\mathscr{N}(n)} g\big( \bd_{t}^{(\ell)} \big) \bigg)^2 \,,
    \label{eq:test_stat_chi_squared}
\end{align}
where $\mathscr{N}(n)$ denotes the set of neighbors of vertex $n$ in the graph $\calG$.
Assuming that $\vect(\bE_t)\sim\calN(\cb{0},\sigma^2\bI)$, where $\vect(\cdot)$ is the vectorization operator, it can be shown using \cite[Proposition~2]{ferrari2019distributed} that $\bd_{t}^{(\ell)}$ is asymptotically distributed as $\bd_{t}^{(\ell)}\stackrel{\tiny a}{\sim} \calN(\cb{0},\eta\sigma^2\bI)$, with
\begin{align}
    \eta = \frac{\lambda}{2-\lambda} + \frac{\Lambda}{2-\Lambda} - \frac{2\lambda\Lambda}{\lambda+\Lambda-\lambda\Lambda} \,.
\end{align}

This means that the graph-filtered $\bd_{t}^{(\ell)}$ is distributed as $g(\bd_{t}^{(\ell)})\stackrel{\tiny a}{\sim}\calN(\cb{0},\bR)$, where 
\begin{align}
    \bR = \eta\sigma^2 \sum_{n=2}^N \big(h(\mu_n)\big)^2 \bu_n\bu_n^\top \,,
\end{align}
for $\ell=1,\ldots,L$.

Similarly, $\sum_{m\in\mathscr{N}(n)} g( \bd_{t}^{(\ell)} )\stackrel{\tiny a}{\sim}\calN(\cb{0},\sigma_R^2(n))$, where
\begin{align}
    \sigma_R^2(n) = \sum_{m,p\in\mathscr{N}(n)} \eta\sigma^2 \sum_{k=2}^N \big(h(\mu_k)\big)^2 \big[\bu_k\bu_k^\top \big]_{m,p} \,,
\end{align}
and $[\,\cdot\,]_{m,p}$ denotes the $(m,p)^{\rm th}$ element of a matrix.

Note that the summands in~\eqref{eq:test_stat_chi_squared} are statistically independent for different values of $\ell$, which means that under the null hypothesis, $\br_{\rm daGFSS}(n)$ follows a Chi-squared distribution with $L$ degrees of freedom.
Then, a p-value of $p_n$ can be obtained by setting the test threshold $\xi_n$ according to
\begin{align}
    \xi_n = \sigma_R(n) \Gamma_{\rm inc}^{-1}\Big(\frac{L}{2}; 1-(1-p_n)\Gamma\Big(\frac{L}{2}\Big) \Big) \,,
\end{align}
where $\Gamma(\cdot)$ is the common gamma function and $\Gamma_{\rm inc}(\cdot\,;\cdot)$ is the upper incomplete gamma function.

Determining whether $t$ is a change point of $\bY_{\!t}$ is now equivalent to testing if a change occurred on at least one of the vertices, $n=1,\ldots,N$, which consists of a multiple testing problem.
In order to limit the probability of type I errors in this context, we employ a False Discovery Rate (FDR) controlling procedure. Since the test statistics $\br_{\rm daGFSS}(n)$ are statistically dependent, we employ the Bonferroni correction as in~\cite{ferrari2019distributed}, since it can be applied efficiently and distributedly in this case. Thus, for an FDR of $\alpha$ we set $p_n=\alpha/N$.

\begin{algorithm} [thb]
\small
\SetKwInOut{Input}{Input}
\SetKwInOut{Output}{Output}
\caption{Remote sensing Graph CD Algorithm~\label{alg:proposed_alg}}
\Input{Images $\{\bY_{\!t}\}$, parameters $0<\lambda<\Lambda<1$, $\gamma>0$.}

Compute the graph $\calG$ based on the first observed HI using the procedure of Section~\ref{sec:sppx_graph_constr} \;
Initialize $\bV_{\!1}=\bY_{\!1}$ and $\bV_{\!1}'=\bY_{\!1}$ \;
\For{$t=2,3,\ldots$}{
Compute $\bV_{\!t}$, $\bV_{\!t}'$ and $\bD_t$ using eqs.~\eqref{eq:newma_adap} and~\eqref{eq:differece_img_newma} \;
Compute the test statistic $\br_{\rm daGFSS}(n)$ using eq.~\eqref{eq:test_stat_chi_squared}\;
\For{$n=1,\ldots,N$}{
\If{$\br_{\rm daGFSS}(n) >\xi_n$}{
Flag $t$ as a change point for vertex $n$ and set $\widehat{c}_{t,n}\leftarrow1$ \;
}
}}
\end{algorithm}


\begin{figure}
    \centering
    \includegraphics[height=0.25\linewidth]{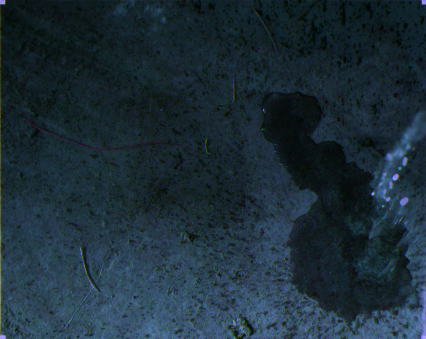} \quad
    \includegraphics[height=0.25\linewidth]{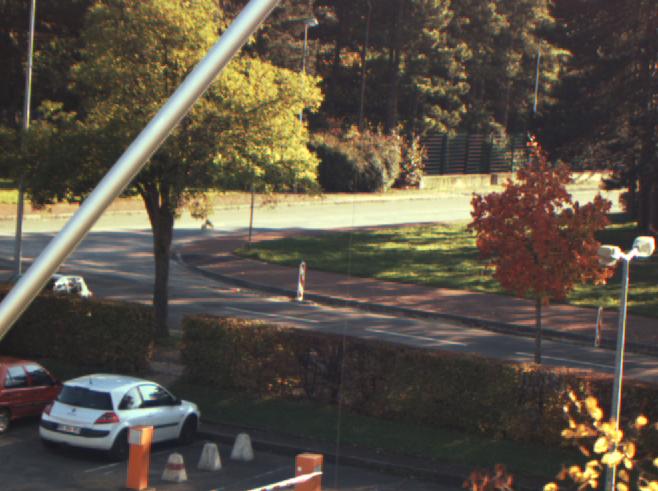}
    \vspace{-0.2cm}
    \caption{Snapshot of the multispectral video sequences used in example~1 (left) and in example~2 (right).}
    \label{fig:snapshots}
\end{figure}

\section{Experimental results} \label{sec:experiments}

In this section, we evaluate the performance of the proposed method using two examples. We compare the proposed method with the traditional change vector analysis (CVA), with MAD~\cite{nielsen1998MAD} and with the IRMAD~\cite{nielsen2007IRMAD} in the first example, and also with the DSFA~\cite{du2019CDdeepSlowFeatureAnalysis} in the second one. 
The first example illustrates the change-point detection performance for an abrupt change, while the second considers a more general scenario where varied amounts of changes are present in the scene. The superpixel decomposition is computed using the SLIC algorithm~\cite{achanta2012slicPAMI}.
Snapshots of the sequences used in both examples are shown in Fig.~\ref{fig:snapshots}.

\begin{figure}
    \centering
    \includegraphics[width=0.48\linewidth]{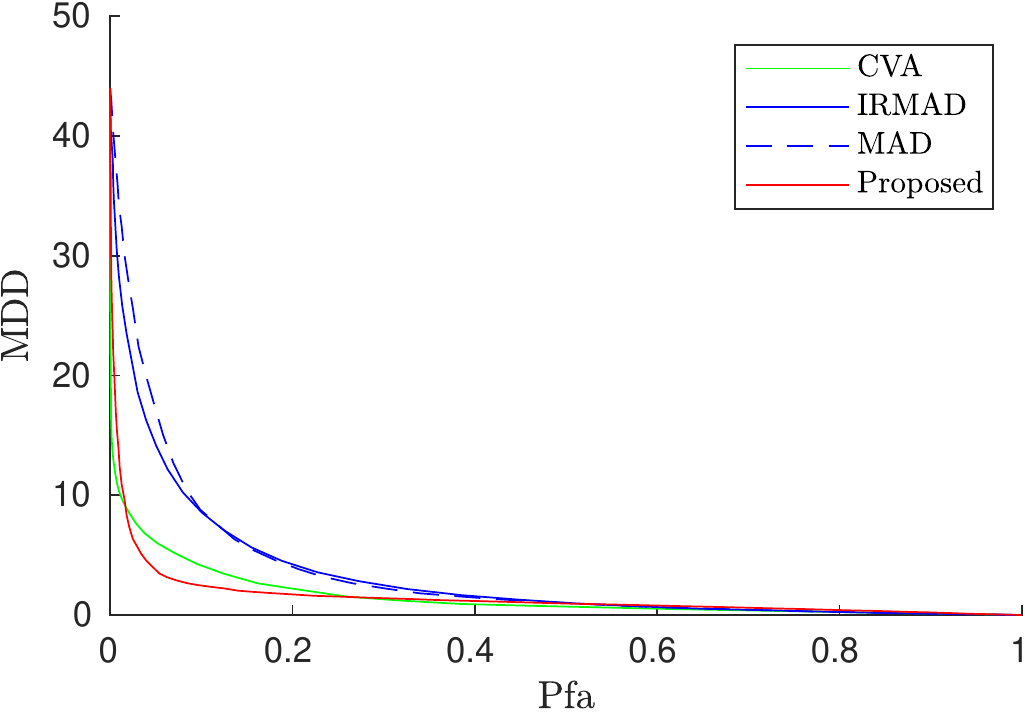}
    \includegraphics[width=0.48\linewidth]{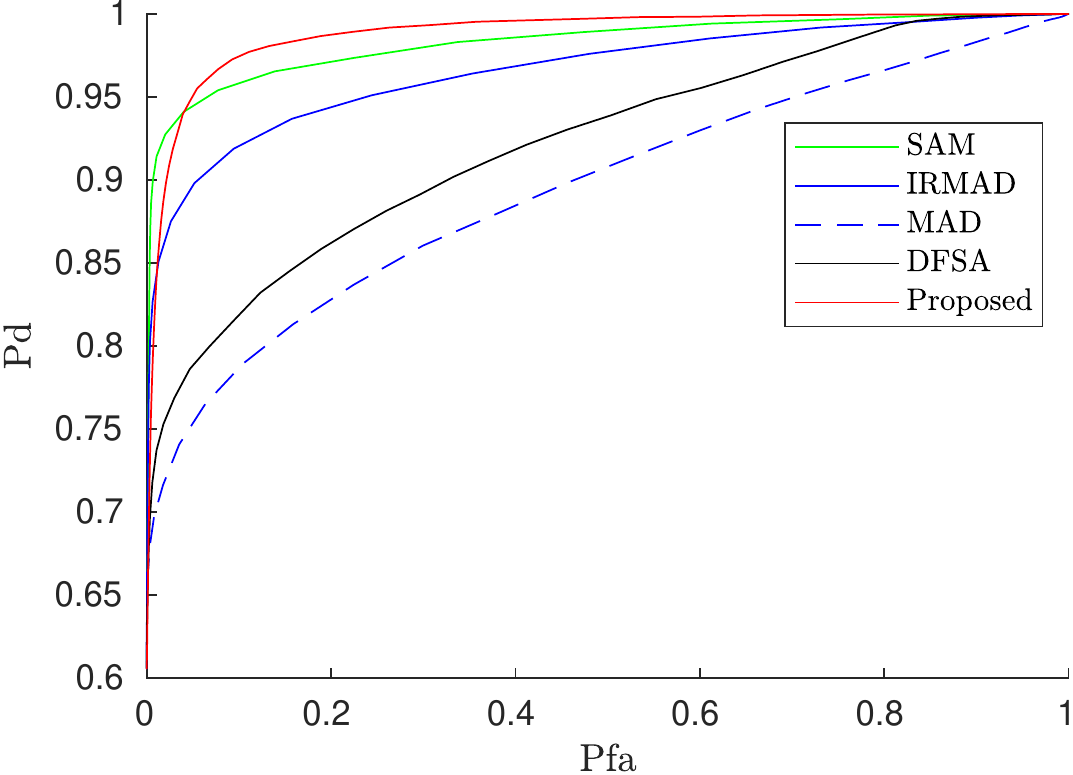}
    \centering (a) \hspace{0.4\linewidth} (b)
    \vspace{-0.3cm}
    \caption{Experimental results. (a) Mean detection delay as a function of the probability of false alarm (Example 1). (b) Probability of detection as a function of the probability of false alarm (Example 2).}
    \label{fig:quantitative_res1}
\end{figure}

\paragraph*{Example 1}

In this example, we consider a video sequence with $T=70$ frames of a concrete floor, which is static at the beginning up until a water bottle is emptied in front of the camera. This generates an abrupt change point at $t_c=16$, after which frequent and intermittent changes are observed until the end of the sequence. 
The images contained $L=9$ bands and were resized to $N=100$ pixels to accelerate processing. Additive white Gaussian noise with a signal to noise ratio of $10$~dB was added to each frame. To compare the performance of the algorithms, we ran a Monte Carlo simulation with $1000$ realizations, and evaluated the mean detection delay ($\Ex\{\widehat{t}_c-t_c\}$, where $\widehat{t}_c$ is the detected change point) as a function of the probability of false alarm (evaluated as the probability of having a detection when $t<t_c$).
The parameters of the proposed algorithm were set as $S=6$, $\gamma=0.1$, $\lambda=0.01$, and $\Lambda=0.8$.
The results are presented in Fig.~\ref{fig:quantitative_res1}-(a), and show that the proposed method is able to provide a detection delay smaller than the other approaches.

\begin{figure}
    \centering
    \includegraphics[width=\linewidth]{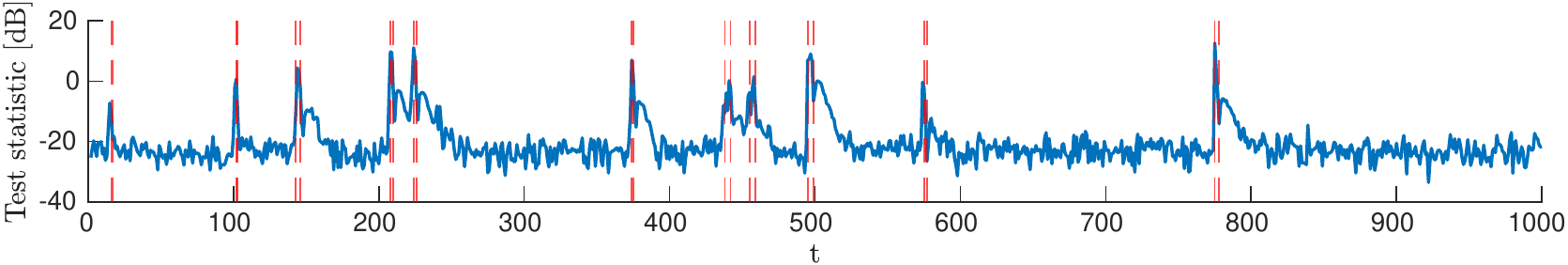}
    \vspace{-0.65cm}
    \caption{Test statistic $\br_{\rm daGFSS}(n)$ in decibels (blue) and true change points (red) for a \emph{road} pixel.}
    \label{fig:testStatisticRoadPx}
\end{figure}

\paragraph*{Example 2}

In this experiment, we consider a real multispectral video sequence of an outdoor scene with intermittent object motion (e.g., passing cars and pedestrians) as well as more subtle, non-informative changes (e.g., tree leaves moving). The video contained $50\times50$ pixels, $L=7$ spectral bands and $T=1000$ frames. Ground truth segmentation results of the moving targets of interest (e.g., cars and pedestrians) were available for this sequence~\cite{benezeth2014MSvideos}, which we use to compute the true detections $c_{t,n}$ and evaluate the detection and localization performance of the algorithms. Denoting the estimated changes by $\widehat{c}_{t,n}$, we compute the probability of detection and false alarm by averaging them across all pixels and across all frames:
\begin{align}
    {\rm Pd} & = \sum_{\forall t,n \,:\, c_{t,n}=1 } \widehat{c}_{t,n} \bigg/ \sum_{\forall t,n} c_{t,n} \,,
    \\
    {\rm Pfa} & = \sum_{\forall t,n \,:\, c_{t,n}=0 } \widehat{c}_{t,n} \bigg/ \bigg(TN - \sum_{\forall t,n} c_{t,n} \bigg) \,.
\end{align}
The parameters of the proposed algorithm were set as $S=5$, $\gamma=0.1$, $\lambda=0.15$, and $\Lambda=0.5$.

The results are shown in Fig.~\ref{fig:quantitative_res1}-(b) and~\ref{fig:testStatisticRoadPx}. It can be seen that even though this sequence is challenging since the changes happen during short and intermittent periods, the proposed algorithm is able to obtain a better performance if the probability of false alarm is not very small. 

\section{Conclusions}

In this paper, we proposed an online graph-based change detection algorithm for multitemporal multiband images. Acting on neighboring spectra as adjacent vertices in a graph, this algorithm focuses on anomalies concurrently activating groups of vertices corresponding to compact, well-connected and spectrally homogeneous image regions. It fully benefits from recent advances in graph signal processing and learns a graph to characterize the support of the data directly from the input image using superpixel decomposition methods. This allows us to exploit both spatial and spectral information. The method is unsupervised as it requires minimal knowledge about the nature of the changes, and is computationally efficient and scalable as it is spatially distributed. Preliminary experiments indicate superior performance of the proposed algorithm both in the detection and localization of changes.

\bibliographystyle{IEEEtran}
\bibliography{references_changeDet_short}

\end{document}